\documentclass[arxiv
,superscriptaddress,twocolumn]{revtex4-1}

\usepackage{lineno,hyperref}
\usepackage{graphicx}
\usepackage{bmpsize}
\usepackage{amsfonts}
\usepackage{soul,color}

\newcommand{\ben}{\begin{eqnarray}}
\newcommand{\een}{\end{eqnarray}}

\newcommand{\bef}{\begin{figure}[h!bt]\centering}
\newcommand{\eef}{\end{figure}}
\newcommand{\bet}{\begin{table}[hbt]\centering}
\newcommand{\eet}{\end{table}}

\begin{document}

\title{A N\'eel-type antiferromagnetic order in the spin 1/2 rare-earth honeycomb YbCl$_3$}
\author{Jie Xing}
\thanks{These authors contributed equally}
\affiliation{Department of Physics and Astronomy and California NanoSystems Institute, University of California, Los Angeles, CA 90095, USA}
\author{Erxi Feng}
\thanks{These authors contributed equally}
\affiliation{Neutron Scattering Division, Oak Ridge National Laboratory, Oak Ridge, Tennessee 37831, USA}
\author{Yaohua Liu}
\affiliation{Neutron Scattering Division, Oak Ridge National Laboratory, Oak Ridge, Tennessee 37831, USA}
\author{Eve Emmanouilidou}
\affiliation{Department of Physics and Astronomy and California NanoSystems Institute, University of California, Los Angeles, CA 90095, USA}
\author{Chaowei Hu}
\affiliation{Department of Physics and Astronomy and California NanoSystems Institute, University of California, Los Angeles, CA 90095, USA}
\author{Jinyu Liu}
\affiliation{Department of Physics and Astronomy and California NanoSystems Institute, University of California, Los Angeles, CA 90095, USA}
\author{David Graf}
\affiliation{National High Magnetic Field Laboratory, 1800 E. Paul Dirac Drive, Tallahassee, FL 32310, USA}
\author{Arthur P. Ramirez}
\affiliation{Department of Physics, University of California, Santa Cruz, CA 95064, USA}
\author{Gang Chen}
\affiliation{Department of Physics and Center of Theoretical and Computational Physics,
The University of Hong Kong, Pokfulam Road, Hong Kong, China}
\affiliation{State Key Laboratory of Surface Physics and Department of Physics, Fudan University, Shanghai 200433, China}
\author{Huibo Cao}
\email{Corresponding author: caoh@ornl.gov}
\affiliation{Neutron Scattering Division, Oak Ridge National Laboratory, Oak Ridge, Tennessee 37831, USA}
\author{Ni Ni}
\email{Corresponding author: nini@physics.ucla.edu}
\affiliation {Department of Physics and Astronomy and California NanoSystems Institute, University of California, Los Angeles, CA 90095, USA}
\date{\today}

\begin{abstract}
Most of the searches for Kitaev materials deal with $4d/5d$ magnets with spin-orbit-coupled ${J=1/2}$ local moments such as iridates and $\alpha$-RuCl$_3$. Here we propose the monoclinic YbCl$_3$ with a Yb$^{3+}$  honeycomb lattice for the exploration of Kiteav physics. We perform thermodynamic, $ac$ susceptibility, angle-dependent magnetic torque and neutron diffraction measurements on YbCl$_3$ single crystal. We find that the Yb$^{3+}$ ion exhibits a Kramers doublet ground state that gives rise to an effective spin ${J_{\text{eff}}=1/2}$ local moment.
The compound exhibits short-range magnetic order below 1.20 K, followed by a long-range N\'eel-type antiferromagnetic order at 0.60 K, below which the ordered Yb$^{3+}$ spins lie in the $ac$ plane with an angle of 16(11)$^{\circ}$ away from the $a$ axis. These orders can be suppressed by in-plane and out-of-plane magnetic fields at around 6 and 10 T, respectively. Moreover, the N\'eel temperature varies non-monotonically under the out-of-plane magnetic fields. The in-plane magnetic anisotropy and the reduced order moment 0.8(1) $\mu_B$ at 0.25 K indicate that YbCl$_3$ could be a two-dimensional spin system to proximate the Kitaev physics.

\end{abstract}

\maketitle

\section{Introduction}
In recent years, there has been a tremendous effort aimed at finding a material that supports a quantum spin liquid (QSL) ground state\cite{QSL1,QSL2}. QSL state has long-range entangled spins, which prevents the breaking of symmetry down to zero temperature. Recently, A. Kitaev proposed a pairwise anisotropic spin model for QSL ground state on a honeycomb lattice \cite{kitaev}, which presents a ${\mathbb Z}_2$ state with gapless and nodal Majorana fermion excitations and gapped bosonic visons. A material realization of the Kitaev model was suggested to be present in honeycomb iridates A$_2$IrO$_3$ (A = Na, Li, H$_3$Li, Cu) and $\alpha$-RuCl$_3$ \cite{AIrO31,AIrO32,AIrO322,AIrO33, AIrO34, AIrO35, AIrO36, AIrO37, AIrO38, AIrO39, Cu, debyetemp2, RuCl31, debyetemp1, RuCl32,rucl3torque, RuCl3magnetization, RuCl34,RuCl35,zigzagRuCl3,RuCl3magstr,rucl3torque2,RuCl33,sears,weiqiang}. The spin-orbit coupling of the iridium or ruthenium moments has been proposed to create highly anisotropic spin interactions including the nearest-neighbor Kitaev interaction \cite{khu}. Due to the extended nature of 4$d$/5$d$ orbits, in A$_2$IrO$_3$ and $\alpha$-RuCl$_3$, in addition to a nearest-neighbor Kitaev interaction, further neighbor interactions often exist, leading to greater complexity. It has been suggested theoretically, however, that rare-earth magnets, especially Yb-based ones may provide a more faithful realization of the Kitaev model~\cite{FYL,rau,motome,REKiteav1,REKiteav2}. 

The rare-earth $4f$ electrons experience much stronger spin-orbit coupling and are more localized comparing to $ 4d/5d $ electrons~\cite{FYL}. The crystal electric field (CEF) enters as a subleading energy scale and splits the spin-orbital coupled $J$ states , often leading a two-fold degenerated ground state, the so-called the effective spin-1/2 (${J_{\text{eff}}=1/2}$), which could present the quantum magnetism in the low dimensional structures\cite{Yb1,Yb3}. Due to the strong localization of the $4f$ electrons, the spin exchange interaction is usually limited to the nearest neighbors. Although the large magnetic moments of rare earth ions can result in strong long range dipole-dipole interaction coupling that exceeds the exchange energy, for Yb$^{3+}$ with $J_{\text{eff}}=1/2$, the dipole-dipole interaction can be ignored as proved in other Yb$^{3+} $~\cite{KateRossPRX, CaoYbTO, PhysRevLett.115.167203}. These properties suggest that Yb-based compounds may be good systems to study the Kitaev model. In this paper, we carry out the first experimental study on the rare-earth honeycomb YbCl$_3$ and map out the magnetic field-temperature ($H-T$) phase diagram. The ${J_{\text{eff}}=1/2}$ magnet YbCl$_3$ exhibits short-range magnetic order (SRO) at 1.20 K and long-range ordered (LRO) N\'eel-type antiferromagnetic state below 0.60 K and strong in-plane magnetic anisotropy. The observations of SRO and LRO, in-plane magnetic anisotropy and reduced order moment demonstrate that YbCl$_3$ is indeed a quasi-two dimensional (2D) frustrated honeycomb that may provide a platform for extending the research of the Kitaev physics.

\section{Experimental methods}
\begin{figure*}
  \centering
  \includegraphics[width=6.8in]{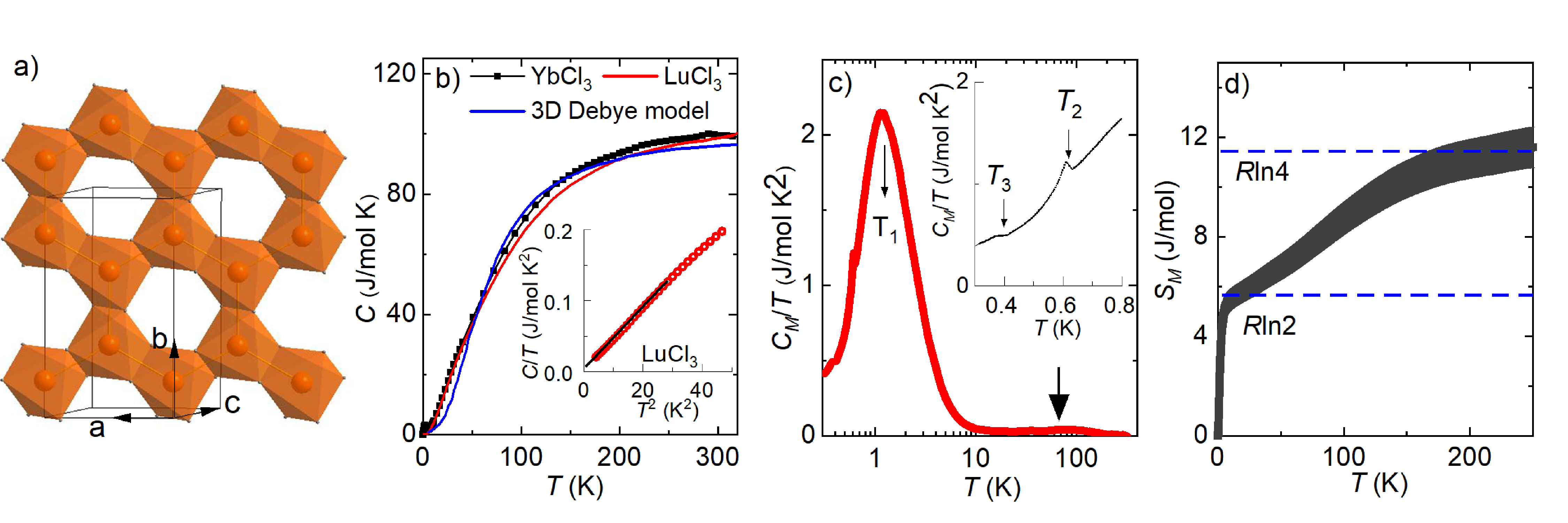}
  \caption{(Color online.) (a) The crystal structure of the $ab$ plane of YbCl$_3$.
  (b) The temperature dependent specific heat $C$ for YbCl$_3$ and LuCl$_3$. The 3D Debye model fitting is shown in blue.
  Inset: $C/T$ vs. $T^2$ for LuCl$_3$. (c) $C_M/T$ vs. $T$ for YbCl$_3$.
  (d) The temperature dependent magnetic entropy $S_M$ with error bars.}
  \label{fig:c}
\end{figure*}

Millimeter-sized transparent YbCl$_3$ single crystals with shiny
as-grown flat $ab$ surfaces were grown by the modified Bridgeman method. Commercial YbCl$_3$ powder (Alfa Aesar 99.99\%) was sealed in a quartz tube under the vacuum and quickly heated up to 800$^{\circ}$C. The ampoule was kept at 800$^{\circ}$C for 10 hours and then cooled to 500$^{\circ}$C at a rate 10$^{\circ}$C/h. The crystals are soft and can be cleaved easily due to its quasi-2D crystal structure. Magnetic susceptibility and specific heat measurements were performed in a Quantum Design VSM Magnetic Property Measurement System (MPMS3) and Physical Property Measurement System (PPMS), respectively. 
The crystals decompose into white powder in air (producing YbCl$_3\cdot$6H$_2$O) within a few minutes, therefore, all sample handling were performed inside a glovebox filled with Ar gas.
Covering the sample with a thin layer of N grease can prevent it from decomposing in air for hours.
During our magnetic measurements, we encapsulated the sample inside a non-magnetic quartz or copper sample can. To avoid the contribution from paramagnetic impurities, the magnetic susceptibility is calculated as  $\chi$=$\Delta M$/$\Delta H = (M(\text{4 T})-M(\text{1 T}))/3$. During our specific heat measurement, we made sure that the sample was fully covered by the N-grease used in the addenda measurement to prevent the sample decomposition and reduce the measurement error.

\begin{table}[b]
\begin{tabular}{p{0.8cm}p{0.8cm}p{1.6cm}p{1.6cm}p{1.6cm}p{1.6cm}}
\hline
\hline
\multicolumn{6}{c}{YbCl$_3$ at 300 K ~~monoclinic \textit{C}2/\textit{m}, 330 reflections }\\\hline
\multicolumn{6}{c}{a=6.730{\AA}~~b=11.5676{\AA}~~c=6.3326{\AA}}\\
\multicolumn{6}{c}{$\alpha$=90.00$^{\circ}\ $ ~~ $\beta$=110.69$^{\circ}$ ~~$\gamma$=90.00$^{\circ}$ }\\
\multicolumn{6}{c}{R$_{F^2}$=0.119 ~wR$_{F^2}$=0.152 ~R$_{F}$=0.115 ~$\chi^2$=4.80}\\
\hline
Atom & Wkf. & x & y & z &U$_{eq}$ \\ \hline
Cl1 & 8j & 0.2594(6) & 0.3204(6) & 0.2402(6) & 0.014(1)\\
Cl2 & 4i & 0.216(1) & 0 & 0.2477(9) & 0.013(2)\\
Yb & 4g & 0 & 0.1675(7) & 0 & 0.011(1)\\
\hline
\hline
\end{tabular}
\caption{The crystal structure of YbCl$_3$ at 300 K.
Wkf. column shows the multiplicity and Wyckoff letter of the site.
U$_{eq}$ is defined as one third of the trace of the U$_{ij}$ matrix
that describes the thermal displacement.}
\label{tab.1}
\end{table}

Single crystal neutron diffraction for YbCl$_3$ was measured on the Four-Circle Diffractometer (HB-3A) at the High Flux Isotope Reactor (HFIR) at Oak Ridge National Laboratory (ORNL) \cite{neutron}. 

\section{Experimental results}
\subsection{Crystal structure of YbCl3}

A good fit to the experimental data suggests the sample is of high quality. The refined crystallographic data
are summarized in Table~\ref{tab.1}. The compound crystallizes in the monoclinic $C2/m$ space group,
the same as $\alpha$-RuCl$_3$ \cite{debyetemp1}. The slightly distorted edge-sharing
YbCl$_6$ octahedra form layered honeycomb $ab$ planes, as shown in Fig.~\ref{fig:c}(a).
The out-of-plane nearest neighbor distance of Yb$^{3+}$ is 6.3326 $\AA$
and the in-plane nearest neighbor distance is 3.90(1) $\AA$ with the ratio of them being 1.62,
slightly less than 1.75 found in $\alpha$-RuCl$_3$ \cite{debyetemp1}.

\subsection{2D honeycomb with ${J_{\text{eff}} = 1/2}$ ground state}
Magnetic specific heat is a powerful tool to identify the ground state
since it provides the entropy release related to possible phase transitions.
The specific heat of YbCl$_3$ and isostructural non-magnetic LuCl$_3$ were
measured at zero magnetic field, as shown in Fig.~\ref{fig:c}(b).
As insulators, their specific heat data can be written as
${C^{\text{YbCl}_3}=C^{\text{YbCl}_3}_{ph}+C_M}$ and ${C^{\text{LuCl}_3}=C^{\text{LuCl}_3}_{ph}}$, where
$C_{ph}$ is the phonon contribution and $C_M$ is the magnetic contribution.
Since both compounds have similar molar mass and crystal structure, ${C^{\text{YbCl}_3}_{ph}= C^{\text{\text{LuCl}}_3}_{ph}}$ to an accuracy of $<$ 1\%. Therefore, we can
isolate the magnetic contribution of YbCl$_3$ by subtracting the lattice part of LuCl$_3$ to obtain $C_M$ by $C_M=C^{\text{YbCl}_3}-C^{\text{LuCl}_3}$, as shown in Fig.~\ref{fig:c}(c).
Upon cooling, unlike $\alpha$-RuCl$_3$ which shows LRO at 7 K via a large lambda anomaly \cite{debyetemp1}, the dominant feature in YbCl$_3$ is a large broad peak centered at around $T_1=1.20$ K, followed by 
a small sharp kink at $T_2=0.60$ K and a subtle feature at $T_3=0.40$ K as shown
in the inset of Fig.~\ref{fig:c}(c). Both $T_1$ and $T_2$ show no sample
variation while $T_3$ is sample dependent and may be from defects/impurities/imperfections. In addition, a second broad weak specific heat hump centered at around 100 K is discernible,
which could be attributed to the Schottky anomaly from the discrete energy levels due to CEF effect.

\begin{figure}
  \centering
  \includegraphics[width=3.5in]{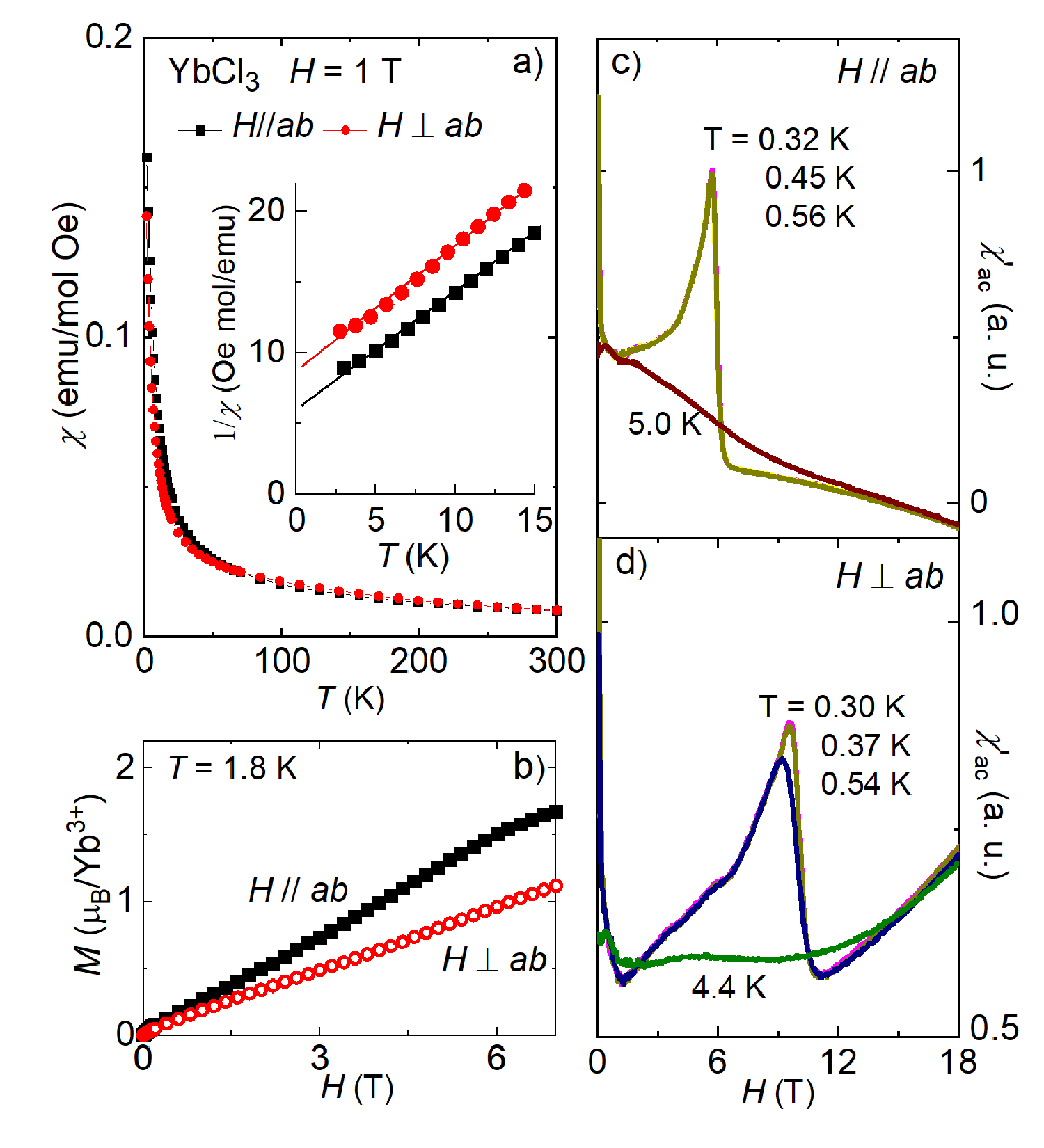}
  \caption{(Color online.)
  (a) The temperature dependent magnetic susceptibility of YbCl$_3$
  at $H=1$ T with ${H \parallel ab}$ and ${H \perp ab}$.
  Inset: the inverse magnetic susceptibility 1/$\chi$ from 3 K to 15 K for ${H \parallel ab}$ and ${H \perp ab}$.
  (b) The isothermal magnetization data taken at 1.8 K with ${H \parallel ab}$ and ${H \perp ab}$.
  (c)-(d): The field dependent $ac$ susceptibility $\chi_{ac}^{\prime}$ with ${H \parallel ab}$ (c) and ${H \perp ab}$ (d) at various temperatures. We used frequency of 577 Hz and current of 0.5 mA.
 }
  \label{fig:cmvst}
\end{figure}

\begin{figure}
  \centering
  \includegraphics[width=3.8in]{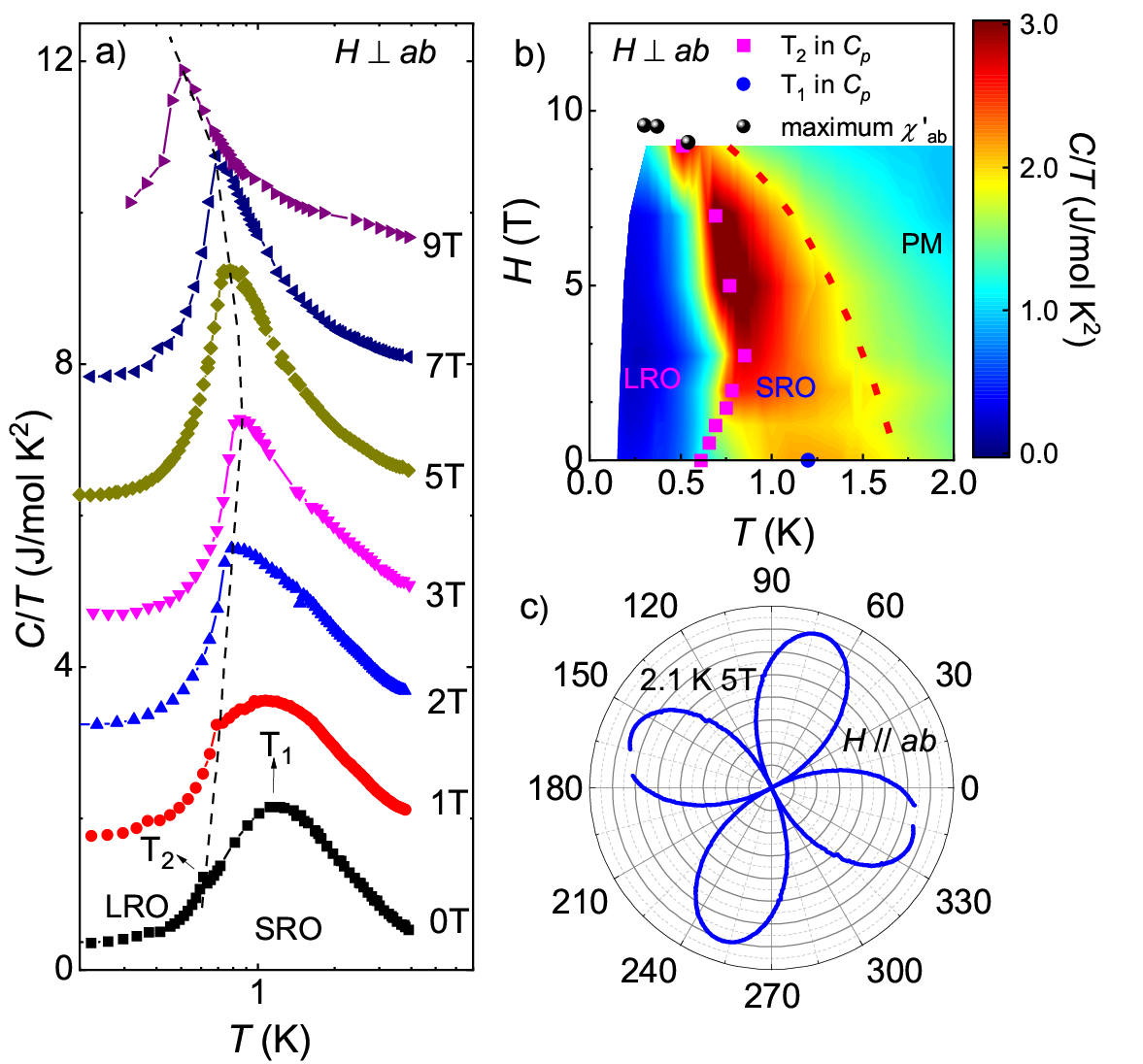}
  \caption{(a) $C/T$ vs. $T$ at various magnetic fields.
    Each data set was offset by 1.5 J/mol K${^2}$. The dash line is a guide to the eyes to track the evolution of the SRO features shown in $C_p$ under field. (b) The $H-T$ Phase diagram of YbCl$_3$. (c) The polar plot of angle-dependent magnetic torque at $T$ = 2.1 K and $H$ = 5 T when the field rotates in the $ab$ plane. The zero-degree-crystal axis $l$ was arbitrarily chosen.}
  \label{fig:torque}
\end{figure}

The magnetic entropy release was calculated based on $S_m=\int C_M/T dT $
and presented in Fig.~\ref{fig:c}(d). It provides information on the CEF
energy splitting scheme. As a Kramers ion, the state $^2F_{7/2}$ of Yb$^{3+}$ would split into four doublets under $C_{2}$ symmetry.
Upon warming, $S_m(T)$ exhibits a
two-plateau feature, suggesting a substantial CEF energy gap between
the ground state and the first excited state. $S_m(T)$ at the first
plateau reaches 5.3(4) J/mol at 8 K, which is very close to $R\ln2$
expected for the paramagnetic state of a spin-1/2 system.
Around 180 K, $S_m(T)$ almost saturates at 11.4(8) J/mol, consistent
with the full magnetic entropy release $R\ln(8/2)$ expected from the Yb$^{3+}$ ion
with doublet ground state \cite{book}. One can estimate the first CEF excited state locating approximately 21 meV, indicating a well isolated ground doublet of Yb$^{3+}$ ion at low temperature.

The inset of Fig.~\ref{fig:c}(b) shows the $C/T$ vs. $T^2$ plot of LuCl$_3$.
By fitting the data from 1.8 K to 6 K with the low temperature limit of
3D Debye model $C=\beta T^3$, we obtain the Debye
temperature as 260(5) K, a little higher than $\sim$ 210 K of $\alpha$-RuCl$_3$ \cite{debyetemp1, debyetemp2}.
Although $C^{\text{LuCl}_3}$ follows the 3D Debye model at low temperatures,
large deviations from the model can be seen in Fig.~\ref{fig:c}(b) at higher temperatures,
suggesting the failure of using this 3D model to describe the phonons here.
This may not be surprising considering that phonons in $\alpha$-RuCl$_3$ above 15 K
can be fitted by 2D Debye model \cite{debyetemp1}.

\subsection{$H-T$ phase diagram of YbCl$_3$}

To further investigate the nature of the anomalies presented in Fig.~\ref{fig:c}(c),
the magnetic susceptibility and specific heat measurements were performed in a magnetic field.
In Fig.~\ref{fig:cmvst}(a) we show the magnetic susceptibility of
YbCl$_3$ measured at 1 T above 1.8 K. No LRO is
observed above 1.8 K. A Curie-Weiss (CW) fit is made using
$1/\chi=C/(T+\Theta_{\text{w}})$, where $\Theta_{\text{w}}$ is the Weiss temperature and $C$ is the Curie constant,
being related to the effective moment $\mu_{\text{eff}}$ by $\mu_{\text{eff}} \approx \sqrt {8C}$.
The fit of the inverse susceptibility from 3 K to 15 K is presented in the inset of Fig.~\ref{fig:cmvst}(a).
The fitted $\Theta_{\text{w}}^{\parallel}$ = -6(1) K, $\Theta_{\text{w}}^{\perp}$ = -9(1) K, $\mu_{\text{eff}}^{\parallel}$ = 3.1(1)$\mu_B$/Yb$^{3+}$ and $\mu_{\text{eff}}^{\perp}$ = 3.0(1)$\mu_B$/Yb$^{3+}$.
The negative $\Theta_{\text{w}}$ values imply the antiferromagnetic in-plane and out-of-plane exchange interactions. The $\Theta_{\text{w}}^{\parallel}$ and $\Theta_{\text{w}}^{\perp}$ are very different, which indicates anisotropic spin interaction as expected for magnetic exchange interaction between Yb$^{3+}$ ions. The inferred $\mu_{\text{eff}}$ is much smaller than 4.54$\mu_B$
of a free $J=7/2$ Yb$^{3+}$ spin, since the Yb$^{3+}$ ions should behave like spin-1/2 ions below 20 K due to the well isolated Kramers doublet ground state. Then one can extract the $g$-factors of in-plane $g^{\parallel} = 3.6(1)$ and out-of-plane $g^{\perp}=3.5(1)$ by using $\mu_{\text{eff}}=g[J_{eff}(J_{eff}+1)]^{1/2}$ and $J_{\text{eff}}=1/2$.

Figure~\ref{fig:cmvst}(b) shows
the isothermal magnetization up to 7 T. No spontaneous magnetism is observed,
again consistent with dominant antiferromagnetic interactions. $M(H)^{\parallel}$ shows a slope change
around 6 T, but remains linear for $M(H)^{\perp}$ up to 7 T. At 7 T, the value of magnetic moment is 1.7 $\mu_B$/Yb$^{3+}$
with $H \parallel ab$ and 1.1 $\mu_B$/Yb$^{3+}$ with $H \perp ab$,
resulting in M$_{\parallel}$/M$_{\perp}\sim 1.5$ at 7 T.

Field-dependent $ac$ susceptibility with ${H \parallel ab}$ and
${H \perp ab}$ were measured and shown in Fig.~\ref{fig:cmvst}(c) and (d). In both directions, a cusp feature is seen at moderate fields, suggesting sharp slope change in $M(H)$. For ${H \parallel ab}$, the feature occurs at around 5.7 T for temperatures below 0.6 K while for ${H \perp ab}$, it appears at around 9.5 T for temperatures below 0.6 K. Combined with the specific heat data under fields (Fig.~\ref{fig:torque}(a)), we will see that the cusp feature is associated with the suppression of LRO and these two fields are near to the critical fields where the LRO is fully suppressed.

In Fig.~\ref{fig:torque}(a) we plot the temperature dependent $C/T$
at various magnetic fields. At zero field, the broad hump centered at $T_1$ releases 99.8\% of the ground state entropy, leaving only 0.2\% for the tiny sharp peak at $T_2$, which is almost 100 orders of magnitude smaller than the entropy release of LRO in $\alpha$-RuCl$_3$ \cite{ruhc}. With increasing fields, the entropy release is suppressed for the transition at $T_1$ but enhanced for the transition at $T_2$. This is a behavior frequently seen under fields for materials with both SRO and LRO, suggesting that the broad hump at $T_1$ is associated with the SRO and the sharp peak at $T_2$ signals LRO.
Furthermore, an unusual response of $T_2$ to the
applied field is observed, as shown in Fig.~\ref{fig:torque}(a). Instead of being monotonically
suppressed by field, $T_2$ first increases from
0.60 K at 0 T to 0.85 K at 3 T and then gets smoothly suppressed
down to 0.50 K at 9 T. This behavior contradicts the mean-field
theory which suggests negative ${\partial T_N}/{\partial H}$ with
field, but rather can be understood when theoretical treatment
beyond the mean-field theory is employed which has shown that
the reduction of spin dimensionality can induce a positive
${\partial T_N}/{\partial H}$ \cite{theory}. The reduction of spin dimensionality
is a small effect leading to a 0.1\% increase of N$\acute{\rm e}$el temperature in 3D magnet, but is larger with reduced dimensionality.
Recently, very similar behavior has been discovered in the entangled 1D
spin chain material, K$_2$PbCu(NO$_2$)$_6$, where  ${\partial T_N}/{\partial H}$
changes from positive to negative with increasing field and a broad specific heat hump associated with SRO is observed at higher temperatures~\cite{art}. 
The reduction of spin dimensionality is suggested to explain the sign change and the broad specific heat hump in K$_2$PbCu(NO$_2$)$_6$. Therefore, this assures that the broad hump at $T_2$ is associated with the SRO and the sharp peak at $T_1$ signals LRO in YbCl$_3$. More importantly, this is the first time that sign change of ${\partial T_N}/{\partial H}$ was observed in a honeycomb material, indicating that YbCl$_3$ is the most 2D honeycomb system until now.

Based on our results in Fig.~\ref{fig:cmvst}(d) and Fig. \ref{fig:torque}(a), we establish
a $H-T$ phase diagram for YbCl$_3$ with $H \perp ab$ using the contour plot, as presented in Fig.~\ref{fig:torque}(b). At 0.60 K, the sample develops LRO. By applying $H\perp ab$, the magnetic field reduces the spin dimensionality, which manifests in the non-monotonic change of $T_2$ with the field. The LRO is expected to be fully suppressed at around 10 T with $H \perp ab$ and at a lower field around 6 T with $H \parallel ab$ as suggested by Fig.~\ref{fig:cmvst}(c). On the other hand, YbCl$_3$ shows SRO centered around $T_1$=1.20 K at zero field. As shown in this contour plot, it is clear to see that the SRO feature moves to lower temperature with increasing field. The dashed line is used to provide a rough guideline to separate the SRO and PM phases since the PM-SRO is a gradual process of building up the spin-spin correlation. 

\begin{figure}
	\centering
	\includegraphics[width=3.4in]{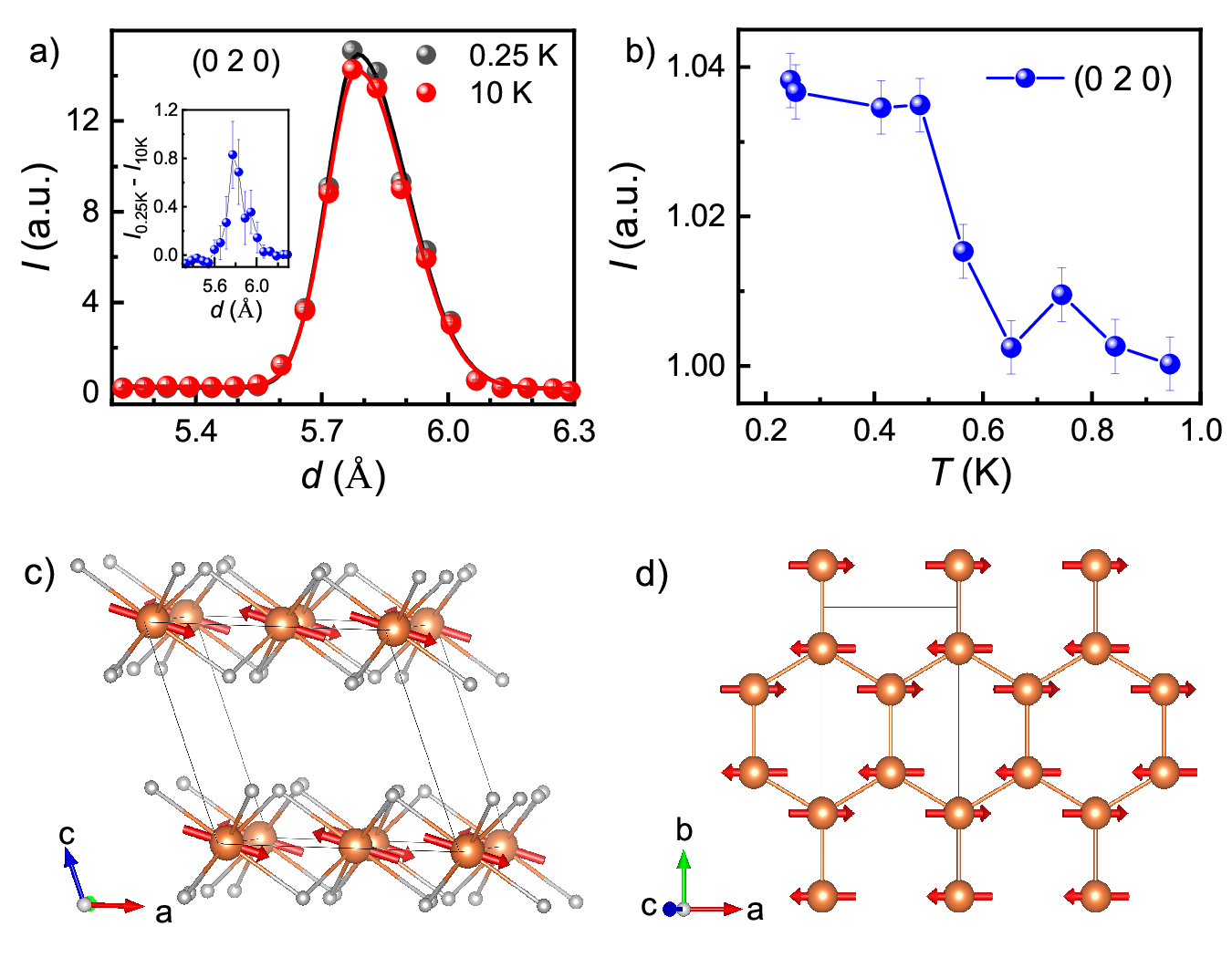}
	\caption{(a) \textit{d}-scan of the peak (0 2 0) in the  monoclinic setting at 0.25 K and 10 K. The inset shows the difference between 10 K and 0.25 K. (b) Temperature dependence of the integrated intensity of the peak (0 2 0). (c-d) N\'eel-type antiferromagnetic structure of Yb$^{3+}$ at 0.25 K, the ordered Yb$^{3+}$ spins lie in the $ac$ plane with the angle between the moment and the $a$ axis as 16(11)$^{\circ}$, nearly parallel to the honeycomb $ab$ plane.} 
	\label{fig:magstr}
\end{figure}

\subsection{N\'eel-type magnetic structure}

Single-crystal neutron-diffraction data were collected at $T=0.25$ K and $T=10$ K on Corelli at SNS (Spallation Neutron Source) at ORNL. No additional Bragg peaks occur at 0.25 K. However, by subtracting the data at 10 K from that at 0.25 K, the difference reveals that the sharp magnetic peaks at the nuclear Bragg positions, indicating the magnetic long-range order with a propagation vector of \textit{k} = 0. The magnetic signal on the top of Bragg peaks (0 2 0) (Fig.~\ref{fig:magstr}(a)) was measured at various temperatures. As shown in Fig.~\ref{fig:magstr}(b), upon cooling, the intensity of the (0 2 0) peak starts to increase at 0.6 K, indicating that LRO emerges below $T_N$ = 0.6 K, consistent with our specific heat results. The magnetic symmetry analysis using MAXMAGN program was employed to solve the magnetic structure \cite{maxmag}. Figure~\ref{fig:magstr}(c-d) present the long-range ordered magnetic structure in YbCl$_3$ determined by single crystal neutron diffraction at 0.25 K. The parent space group $C$2/$m$ with the $k$ vector allows four possible maximal magnetic space groups. The $C$2$’$/$m$ (\#12.60), corresponding to the N\'eel-type antiferromagnetic order in which the spins lie in $ ac $ plane and are stacked in parallel along $c$, is the only one that fits the observed magnetic peaks. The angle between the ordered moment and $a$-axis is 16(11)$^{\circ}$, making the spins tilting toward the $ab$ plane. This magnetic structure supports the $M(H)$ data plotted in Fig. 2(b), which indicates the ordered spins enter the forced ferromagnetic state at a lower field with ${H \parallel ab}$. The refinement of the obtained magnetic peaks gives rise to the ordered Yb$^{3+}$ magnetic moment of 0.8(1) $\mu_B$ in the $ac$-plane. This value is around one third of the fully-ordered moment of 2.24 $\mu_B$/Yb$^{3+}$ expected from the CEF ground doublet ~\cite{CEFYbCl3}, implying strong quantum fluctuation exists at 0.25 K.

\subsection{Anisotropic in-plane bond-dependent coupling}

Since the Kitaev model describes a spin 1/2 honeycomb lattice with highly anisotropic
couplings between nearest neighbors, to obtain some information of the nearest neighbor
coupling, we investigated the in-plane magnetic anisotropy by measuring the angular
dependence of the magnetic torque on the YbCl$_3$ single crystal with $H \parallel ab$ using a cantilever. The data taken at 2.1 K and 5 T are depicted in Fig.~\ref{fig:torque}. $\theta$ is the angle between $H$ and the arbitrarily chosen crystal axis $l$ in the $ab$ plane.
The magnetic torque is very sensitive to the anisotropy of the in-plane magnetization \cite{rucl3torque,rucl3torque2}.
For the Yb$^{3+}$ ion, a large portion of the local moment comes from the orbital
degrees of freedom. Because the orbitals have orientation, the spin-orbit-coupled
local moment would inherit the orbital orientation, and thus the interaction between
the local moment would have a strong orientation dependence (or equivalently, bond orientation dependence)~\cite{WWW}.
Similar to the pyrochlore magnet Yb$_2$Ti$_2$O$_7$~\cite{CaoYbTO,KateRossPRX}, the strong spin-orbital entanglement in the honeycomb magnet YbCl$_3$ brings the strong bond dependent interaction. The bond dependent interaction can be determined by the lattice symmetry (or space group symmetry) and is a reflection of the lattice symmetry. For instance, the four-fold symmetry in the magnetic torque was observed in $\alpha$-RuCl$_3$ above or below the zig-zag LRO, suggesting the bond dependent exchange interactions \cite{rucl3torque,rucl3torque2}. 
As shown in Fig.~\ref{fig:torque}(c), this bond dependent anisotropy is readily manifested
in the magnetic torque measurement. Under magnetic field, the magnetic torque indeed shows four-fold symmetry which agrees with the monoclinic structure and implies the existence of the bond-dependent exchange interactions in YbCl$_3$. 

\section{discussion}
We have established that the ground state
of layered YbCl$_3$ has a 2D honeycomb spin lattice of $J_{\text{eff}}=1/2$ Yb$^{3+}$ spins and revealed the N\'eel-type magnetic structure with reduced moment and anisotropic in-plane bond-dependent coupling, satisfying the prerequisites of the Kitaev model. The N\'eel-type antiferromagnetic order with reduced moments makes YbCl$_3$ honeycomb distinct from the well-studied $4d$/$5d$ honeycomb lattice Na$_2$IrO$_3$ \cite{AIrO322} and RuCl$_3$ \cite{debyetemp1,RuCl3magstr,zigzagRuCl3} hosting the zig-zag magnetic order. In the phase diagram of the nearest-neighbor Heisenberg-Kitaev model \cite{revJPCM2019}, the honeycomb lattice exhibits a zig-zag magnetic order in the region with the ferromagnetic Heisenberg interaction and a N\'eel-type order for antiferromagnetic Heisenberg interaction. Referring to the phase diagram of the Kitaev-Heisenberg model~\cite{KHModel}, a magnetic ground state in YbCl$_3$ maybe lies next to the Kitaev spin liquid from the antiferromagnetic Heisenberg side~\cite{revJPCM2019}. Further investigations, such as inelastic neutron scattering is needed to settle down this problem.

\section{Conclusion}

In conclusion, we propose YbCl$_3$ as a 2D Kitaev material candidate with $J_{\text{eff}}=1/2$ local moments and
strong in-plane magnetic anisotropy. This compound exhibits SRO peak at 1.20 K and LRO below $T_N$=0.60 K. The application of external magnetic fields can suppress these orders at around 6 T (in-plane field) and 10 T (out-of-plane field). The in-plane magnetic anisotropy and the N\'eel-type magnetic order with reduced order moment 0.8(1) $\mu_B$ at 0.25 K suggest that YbCl$_3$ could be a 2D honeycomb to proximate the Kitaev physics.

\section{Acknowledgments}

Ni thanks Prof. Jeffrey G. Rau for the inspiration to work on this material. Work at UCLA was supported by NSF DMREF program under the award NSF DMREF project DMREF-1629457. Work at UCSC was supported by DOE grant DE-SC0017862. Work at ORNL was supported by US DOE BES Early Career Award KC0402010 under Contract DE-AC05-00OR22725 and used resources at the Spallation Neutron Source and the High Flux Isotope Reactor, DOE Office of Science User Facilities operated by the Oak Ridge National Laboratory. Work at NHMFL, Tallahassee is supported by NSF
through NSF/DMR-1644779 and the State of Florida. GC acknowledges the support by
the ministry of science and technology of China with Grant from No.~2016YFA0301001
and 2016YFA0300500.

\appendix

\section{Magnetic symmetry analysis}

\begin{figure}
	\centering
	\includegraphics[width=3.4in]{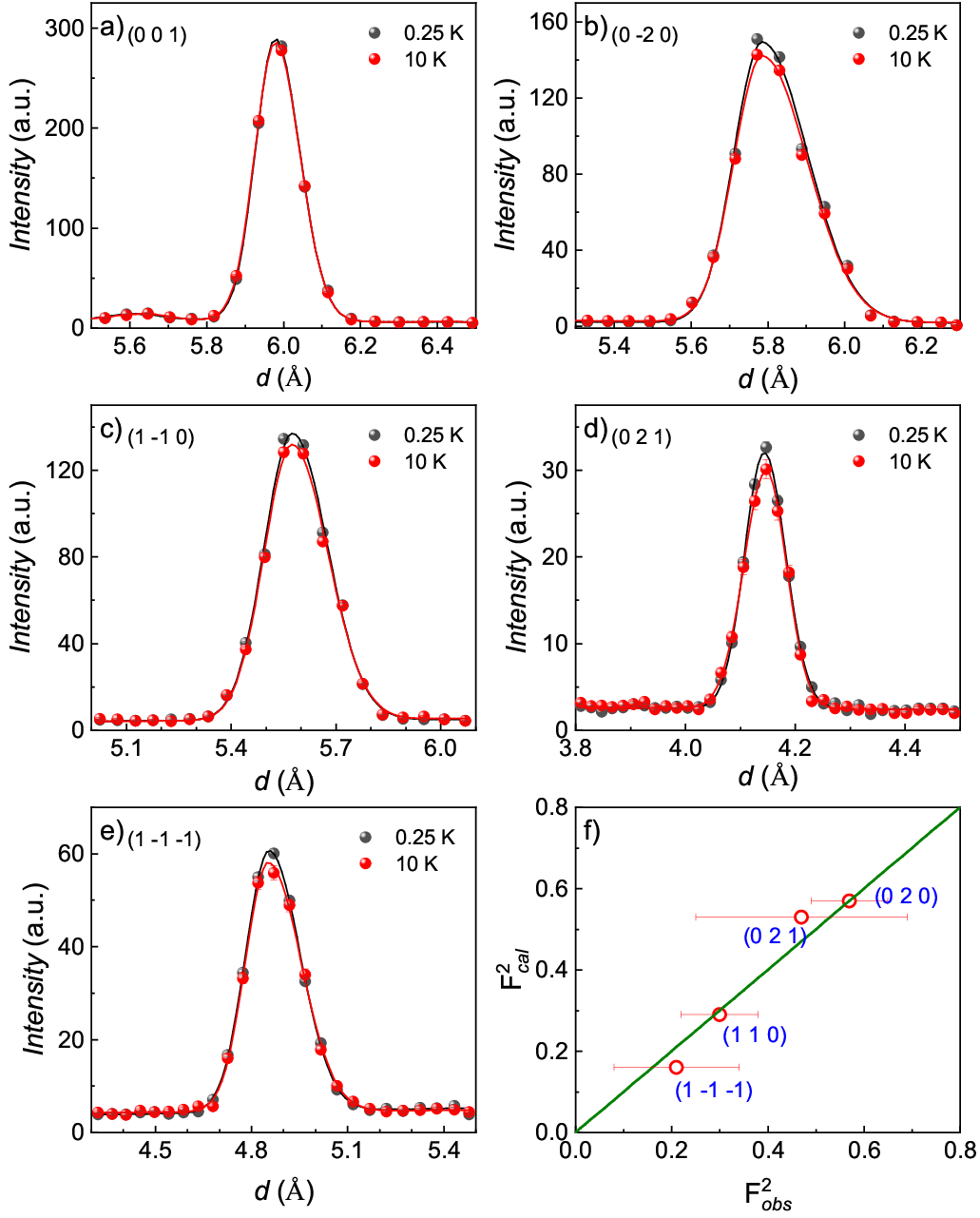}
	\caption{(a-e)The $d$-scan of the peak (0 0 1), (0 -2 0), (1 -1 0), (0 2 1) and (1 -1 -1) at 0.25 K and 10 K. Most of the error bars are smaller than the size of spheres. The solid line represents the fitting of the peaks by a bi-gaussian peak function. (f) Comparison of the squared magnetic structure factors between observations and calculations. The magnetic intensities have been normalized to their nuclear intensities.}
	\label{fig:magpeak}
\end{figure}

Magnetic structure models compatible with the parent space group $C2/m$ and the propagation vector \textbf{k} = (0, 0, 0) are explored using the magnetic symmetry approach using the MAXMAGN program \cite{MAXAGN}. There are four possible maximal magnetic space groups as listed in Table SII. All the four models are colinear magnetic structure and ferromagnetically aligned between hexagonal layers. The moments are either along b axis or lying in ac plane. Fig.~\ref{fig:magpeak} present the d-scan of peak (0 0 1), (0 2 0), (1 1 0), (1 -1 -1) and (0 2 1) at 0.25 K and 10 K. The $C2'/m'$ (\# 12.62) and $C2/m$ are corresponding to the ferromagnetic orders which can be immediately ruled out due to the absence of the magnetic signal at (0 0 1). The $C2/m'$ (\# 12.61) is the N\'eel-type AFM order with magnetic moments along b axis, however, the magnetic peak (0 2 0) is forbidden in the MSG. Thus, only the $C2'/m$ (\# 12.60), which is also N\'eel-type but moment in the ac plane, is suitable for our observations. 

To refine the magnetic structure in Fullprof suite\cite{Fullprof}, the self-calibration of the magnetic peak intensity was performed as following: 

\begin{equation}
F^2_{{hkl}, mag}=\frac{I^{0.25 K}_{hkl}-I^{10 K}_{hkl}}{I^{10 K}_{hkl}}F^2_{{hkl}, nuc}
\end{equation}

where $I^{0.25 K}_{hkl}$ and $I^{10 K}_{hkl}$ are the integrated intensities of peak $hkl$ at 0.25 K and 10 K, obtained in d-space as shown in Fig.~\ref{fig:magpeak}. $F^2_{{hkl}, nuc}$ is the squared nuclear structure factor calculated in Fullprof. $F^2_{{hkl}, mag}$ is the self-calibrated magnetic peak intensity. In total 14 peaks (2 nuclear and 12 magnetic) were extracted, and self-calibrated. Then, they were merged into 5 non-equivalent reflections and used to refine the magnetic structure in Fullprof.

\begin{table*}
\begin{tabular}{cccccc}
\hline
Magnetic & Observation  & C2'/m' (12.62)  & C2/m' (12.61)  & C2'/m (12.60) & C2/m (12.58)\\
reflections&~&FM order&N\'eel type&N\'eel-type&FM order\\
~&~ &in a-c plane&along b axis&in a-c plane&along b axis\\
\hline
(0 0 1)	& -	& $\bullet$ & -&-&$\bullet$ \\\hline
(0 2 0)	&  $\bullet$	& $\bullet$ & -& $\bullet$ &-\\\hline
(1 1 0)&  $\bullet$&  $\bullet$&  $\bullet$&  $\bullet$&  $\bullet$\\
\hline
\end{tabular}
\caption{Magnetic space group (MGS) and the allowed magnetic reflections. $\bullet$ and - represent the allowed and forbidden magnetic reflections in the corresponding magnetic space group, respectively. }
\label{tab.2}
\end{table*}

\end{document}